\documentclass[graybox,envcountchap,sectrefs]{usb_TEX}

\usepackage{hyperref}
\hypersetup{
    pdfauthor={Wilder Castellanos},
    pdftitle={Internet of things: a multiprotocol gateway as solution of the interoperability problem},
    colorlinks,
    citecolor=blue,
    filecolor=black,
    linkcolor=blue,
    urlcolor=blue
}
\usepackage{subfigure}
\usepackage{mathptmx}
\usepackage{helvet}
\usepackage{courier}
\usepackage{apacite}
\usepackage{type1cm}         
\usepackage{makeidx}         % allows index generation
\usepackage{graphicx}        % standard LaTeX graphics tool
\graphicspath{{Figures/}}
\usepackage{multicol}        % used for the two-column index
\usepackage[bottom]{footmisc}% places footnotes at page bottom
\usepackage{iftex}
\usepackage{amsmath}
\usepackage{amssymb}
\usepackage{captdef}
\usepackage{chemformula}
\usepackage{multirow}
\makeindex             % used for the subject index
\newcommand*{\doi}[1]{\href{https://doi.org/#1}{#1}}

\newcommand{\chapterauthor}[1]{%
  {\vspace*{-70pt}%
   \linespread{1.1}\large\scshape#1%
  \vspace*{35pt}}
}
\makeatother

\begin{document}

\mainmatter%%%%%%%%%%%%%%%%%%%%%%%%%%%%%%%%%%%%%%%%%%%%%%%%%%%%%%%

%%%%%%%%%%%%%%%%%%%%% chapter.tex %%%%%%%%%%%%%%%%%%%%%%%%%%%%%%%%%
\chapter*{Chapter 4. Advanced techniques for adaptive video streaming in SDNs}
\chapterauthor{Wilder Castellanos\footnote{Programa de Ingenier\'ia de Electr\'onica, Universidad de San Buenaventura sede Bogot\'a, Cra. 8H No. 172-20, Bogot\'a, Colombia. e-mail: wcastellanos@usbbog.edu.co}, Juan Gabriel Bustos\footnote{Programa de Ingenier\'ia de Electr\'onica, Universidad de San Buenaventura sede Bogot\'a, Cra. 8H No. 172-20, Bogot\'a, Colombia. e-mail: jbustos@academia.usbbog.edu.co} and Fran Wilson Sanabria\footnote{Programa de Ingenier\'ia de Electr\'onica, Universidad de San Buenaventura sede Bogot\'a, Cra. 8H No. 172-20, Bogot\'a, Colombia. e-mail: sfran@academia.usbbog.edu.co}}

%\chapauthor{Wilder Castellanos\footnote{Programa de Ingenier\'ia de Electr\'onica, Universidad de San Buenaventura sede Bogot\'a, Cra. 8H No. 172-20, Bogot\'a, Colombia. e-mail: wcastellanos@usbbog.edu.co}, Juan Gabriel Bustos\footnote{Programa de Ingenier\'ia de Electr\'onica, Universidad de San Buenaventura sede Bogot\'a, Cra. 8H No. 172-20, Bogot\'a, Colombia. e-mail: jbustos@academia.usbbog.edu.co} and Fran Wilson Sanabria\footnote{Programa de Ingenier\'ia de Electr\'onica, Universidad de San Buenaventura sede Bogot\'a, Cra. 8H No. 172-20, Bogot\'a, Colombia. e-mail: sfran@academia.usbbog.edu.co}}

\label{Chapter_4} % Always give a unique label
% use \chaptermark{}
% to alter or adjust the chapter heading in the running head
\abstract{
In the coming years a new management model of telecommunication networks will begin to be implemented. This model, known as Software Defined Networks, implies a radical change in the way networks are conceived today. Therefore, it is necessary to develop studies that allow to know this scenario and its implications during the transmission of different types of traffic. Especially, it is important to analyze the behavior of video traffic due to the significant increase of these flows through networks as well as to the massive use of new video services with ultra-high resolution formats. This fact also implies an enormous consumption of network resources like bandwidth. Likewise, the likelihood to impact on other traffic flows that coexist in the same network is increased caused by network congestion. Under these conditions the ability to adapt the video streams to the available resources of the network is an essential requirement in modern video services in order to prevent network congestion. Therefore, adaptive techniques for video transmission have become an important area of interest for researchers, technology providers and network operators.\newline\indent
This chapter briefly describes a study on software-defined networks and the tools necessary for video transmission on this type of networks. Among the aspects presented is the methodology used to establish the video transmission and the procedure to evaluate the quality obtained. A video transmission experiment is presented, which consists on the evaluation of adaptive transmission of video streams using adaptive techniques as MPEG/DASH and scalable coding. The experiments were carried out over a software-defined network topology in the MININET platform.
\newline\indent
\textit{Keywords:} Multimedia Delivery, Adaptive Video Streaming,  Software Defined Networks, DASH, SHVC}

\section{Introduction}
\label{sec:4.1}
With the inevitable implementation of SDNs (Software Defined Networks) by telecommunication network operators, it is necessary to start the evaluation of the multimedia flow transmission on this scenario. The new paradigm that involves SDNs includes the relocation of the control plane, currently located in network devices (routers and switches), to a single external element, the controller \cite{onf_2012}. This implies that instead of using distributed protocols such as BGP or OSPF, the controller is the software component responsible for routing and network control using standard management software.

In this new model for future networks and taking into account that video traffic is the predominant traffic through current networks, it is essential to develop experiments that study the optimization of the transmission of these flows so that increasing demand for video services can be satisfied, using network resources as efficiently as possible, without sacrificing the user quality experience.

Several aspects are relevant in this area, but mainly those that are related to providing users with video services and the video quality they demand, despite many times their connections do not have good communication quality management. In such scenarios, ability to adapt the video streams is essential. In this sense, multiple solutions have been proposed in recent years, such as more efficient encoders that allow to obtain high quality videos requiring less data.

This chapter describes the main concepts of software-defined networks, video coding and adaptive video transmission. In addition, we present a study where two techniques for adaptive transmission were analyzed: one corresponds to the use of MPEG/DASH (Dynamic Adaptive Streaming over HTTP) as an adaptation mechanism and the second consists in the use of the scalable coding standard SHVC (Scalable High Efficiency Video Coding). Also, we propose a methodology for the development of adaptive video transmission analysis on the MININET emulation platform \cite{lantz_2017}.

This chapter is organized as follows. Section 2 gives a brief forecast overview about the video services and the traffic problem in current networks. In Section 3 we give a brief introduction to the reference model for software defined networks. Section 4 summarizes the basic concepts on adaptive video streaming and then provides an overview on SHVC and DASH standards. The methodology and results of the performance evaluation of the adaptive methods are presented in Section 5. Finally, conclusions are presented in Section 6.

\section{Video Streaming: the digital revolution}
\label{sec:4.2}

The growing demand for streaming video services has resulted in a considerable increase in traffic through data networks. Some studies estimate that by 2022 82\% of network traffic will correspond to video traffic (see Fig.~\ref{ch4_fig:1}, which would be equivalent to every second, one million minutes of video being transmitted over the network \cite{Cisco_2019}. This figure includes all possible forms of IP video which includes Internet video, IP VoD, video conferencing, video-streamed gaming and video files exchanged through file sharing.  

\begin{center}
\includegraphics[scale=1.0]{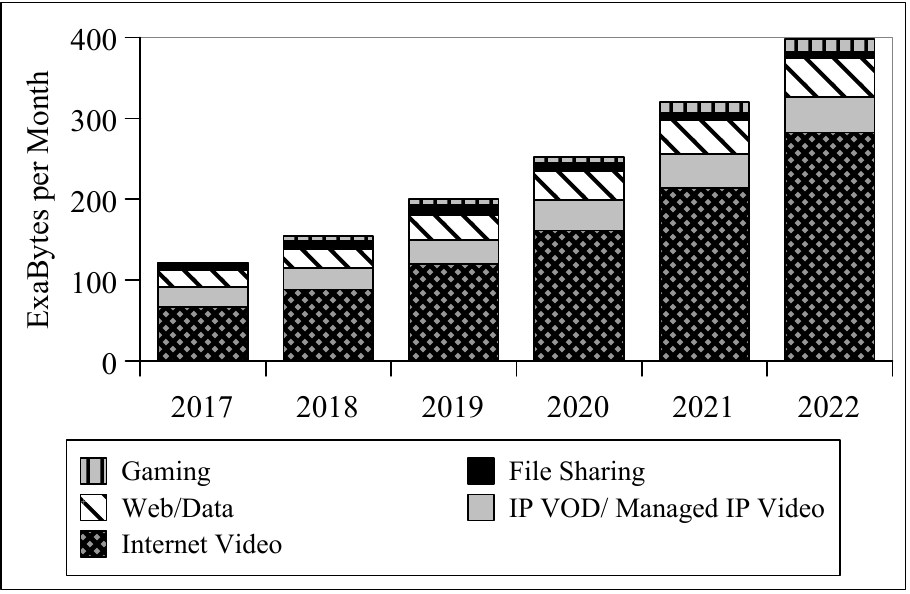}
\figcaption{Global IP traffic by application category \protect\cite{Cisco_2019}}
\label{ch4_fig:1}       % Give a unique label
\end{center}

%\begin{figure}[h!]
%\centering
%\includegraphics[scale=1.0]{Figures/ch_4_fig1_iptraffic.eps}
%\caption{Global IP traffic by application category \cite{Cisco_2019}}
%\label{ch4_fig:1}       % Give a unique label
%\end{figure}

Ericsson also forecasts a dramatic increase in video traffic in its “Mobility Report” \cite{ericsson2017}. This study foresees an increase in video traffic from mobile devices, from 16.59 ExaBytes/month (2018) to 97 ExaBytes/month in 2024. This means a 34 percent compound annual growth rate (CAGR). Globally, the total of video traffic in Internet will grow from 66.9 EB per month in 2017 to 282.3EB in 2022 (33\% CAGR) \cite{Cisco_2019}. This traffic represents 1.1 million of minutes of video transmitted every second.

This significant increase in video traffic is caused by new habits in the consumption of audiovisual content, since it is increasingly common to watch video contents in streaming mode, mainly due to the massive delivery of video on Internet and services such as Netflix, Hulu and YouTube. But not only streaming video services attract the attention of network operators.Video traffic growth is also driven mainly by: the increase of embedded video in many online applications, growth of video-on-demand (VoD) streaming services in terms of both subscribers and viewing time per subscriber, the evolving devices towards larger screens and higher resolutions and the new evolved 4G deployments that increase network performance. This fact is causing a change in the pattern of internet traffic that is going from being a traffic relatively steady to a more dynamic and bursty pattern traffic.

In particular, live Internet video has the potential to generate large amounts of traffic since it replaces the consumption of hours of traditional broadcast content. Live video already represents a 5 percent of Internet video traffic and will increase to reach 17 percent by 2022 \cite{Cisco_2019}. This means an increase of 15-fold in 5 years (72.7\% CAGR). Likewise, video surveillance traffic is of a very different nature than on-demand or live video and represents an upstream video traffic, uploaded continuously to the Internet. Fig.~\ref{ch4_fig:2} shows the growth of video traffic by subcategory. It is important to note that, long-form video will be 175 EB/month in 2022, 82.3\% of all Internet video traffic (including live and short-form video).

\begin{center}
\includegraphics[scale=0.83]{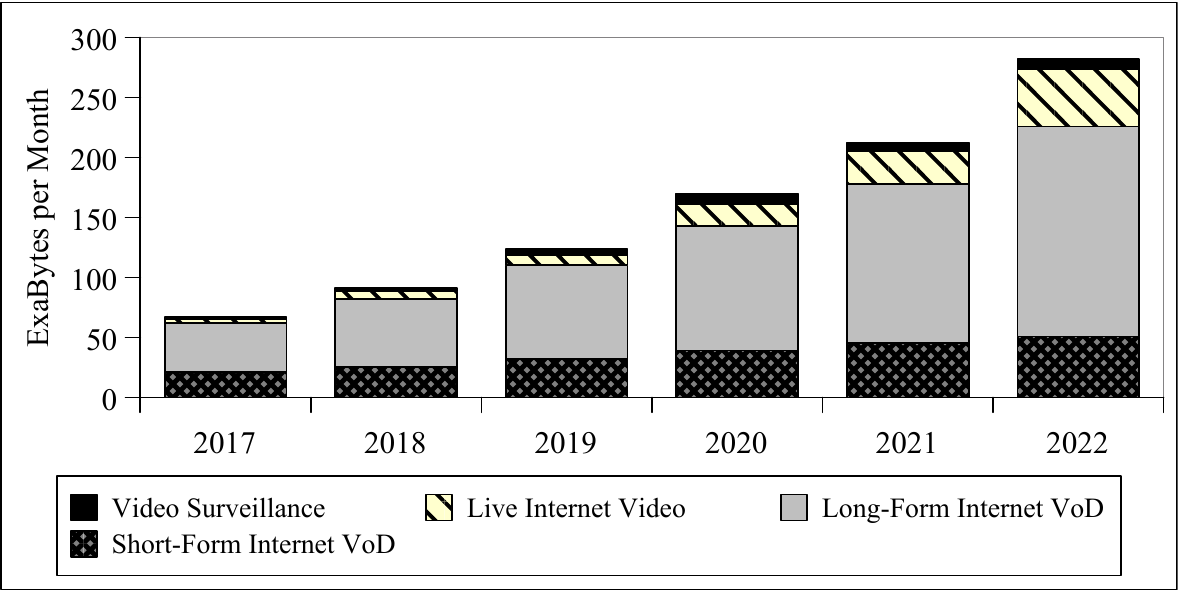}
\figcaption{Internet Video Streaming by Service Type \protect\cite{Cisco_2019}}
\label{ch4_fig:2}       % Give a unique label
\end{center} 
 
%\begin{figure}[h!]
%\centering
%\includegraphics[scale=1.0]{Figures/ch_4_fig_2_Internet_Video_Streaming.eps}
%\caption{Internet Video Streaming by Service Type \cite{Cisco_2019}}
%\label{ch4_fig:2}       % Give a unique label
%\end{figure}

With the arrival of the new generation wireless architectures, as well as the development of cameras and smartphones with the ability to record and play videos with ultra high definition (UHD) resolutions, user expectations are high as regards the possibility of transmitting higher quality video. This directly impact the network capacity since bit rate for a UHD video is about 18 Mbps, more than double the HD video bit rate and nine times more than Standard Definition (SD) video bit rate. And, UHD video will be 22\% of global Internet video traffic by 2022 (see Fig.~\ref{ch4_fig:3}). A trend that will increase with the arrival of 5G since users also expect immersive media and applications as well as 360-degree video and virtual/augmented reality \cite{mattos_2016}. 

\begin{center}
\includegraphics[scale=0.9]{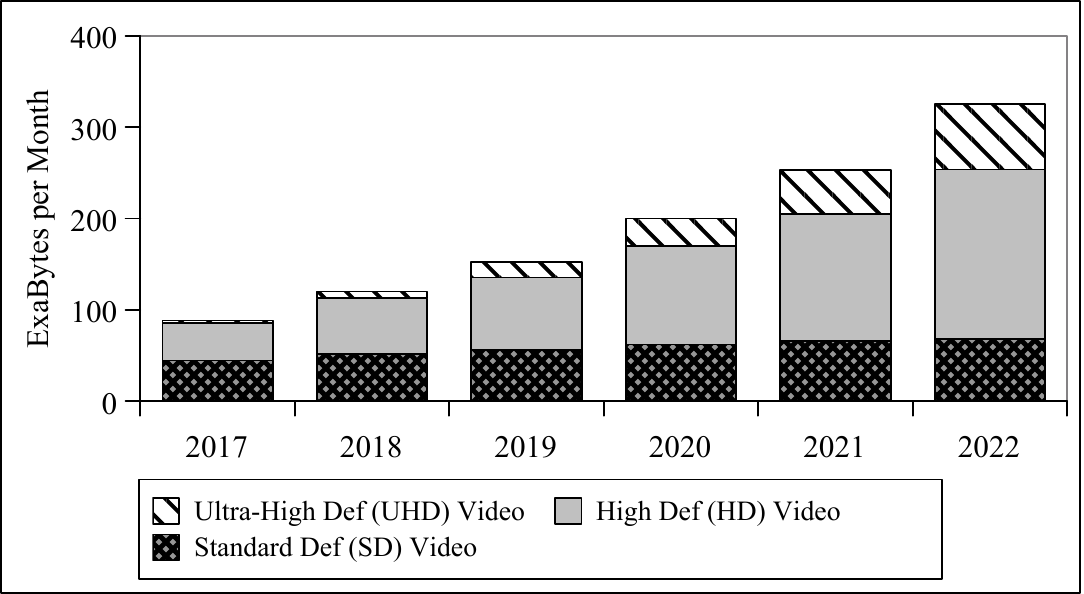}
\figcaption{Global UHD IP video traffic \protect\cite{Cisco_2019}}
\label{ch4_fig:3}       % Give a unique label
\end{center}

%\begin{figure}[h!]
%\centering
%\includegraphics[scale=1.0]{Figures/ch_4_fig_3_Global_UHD.eps}
%\caption{Global UHD IP video traffic \cite{Cisco_2019}}
%\label{ch4_fig:3}       % Give a unique label
%\end{figure}

This implies that efforts should be focused on the development and evaluation of new coding and compression methods, to produce videos that use the least amount of resources during the transmission and/or storage of that content.

Regarding video encoders, the most recently developed is the so-called H.265 / HEVC (High Efficiency Video Coding). This new standard promises to offer a 50\% better compression compared to the predecessor standard H.264, keeping the video with a reduced loss of quality. However, a greater complexity in compression algorithms is required  \cite{ye-kui_2016}. The H.265 Standard is a technique that allows you to respond to the challenge of optimizing UHD video transmission. This standard also has an extension for scalable video encoding SHVC (Scalable High Efficiency Video Coding), which promises to be a possible solution to UHD video transmission problems, providing adaptability in different topologies networks. Basically, SHVC encodes the video in several layers. This scalability feature allows the video to be adapted by transmitting only those layers that can be supported by network according to the available bandwidth estimation. This adaptation thus prevents network congestion and maintains a good level of video quality. In the same way, there is a technique called DASH (Dynamic Adaptive Streaming over HTTP) which consists of encoding a video with different qualities obtaining several representations of the same video. Each representation is divided into segments of short duration. A DASH Server will try to transmit those segments belonging to the representations of highest quality that can be supported by network.

\section{Software Define Networks}
\label{sec:4.3}

Today communication networks have increased their complexity caused by several reasons. Mainly due to the characteristics of new services and new consumption habits of users. For them mobility, security and a connection everywhere and every time are the priorities. It is difficult to overcome these challenges with the management model of the current networks. This is evidenced when services running on the network must be interrupted as soon as new services need to be implemented. In addition to the amazing increase in traffic from emergent services such as online games and virtual reality, new problems of adaptability and scalability are emerging. This has highlighted the need for a new management model for conventional networks.

Software Defined Networks involve a new vision on the management of communication networks. The main feature is the transfer of the control plane of physical devices to a new device, called a controller. This controller is a new network device in charge of manage or control a whole network. This fact allows network manager to configure devices such as switches, routers, network translators (NATs) or firewalls, using centralized software. In SDN model all networks devices can be programmed directly from the controller. In addition, more flexibility is obtained since network administrators can dynamically adjust traffic flows, in order to meet the changing needs of the services. Moreover, network managers can configure, manage and optimize network resources very quickly through dynamic and automated programs, which can be written by themselves. Since open standards are used, Software Defined Networks simplify the design and operation tasks of the networks, because instructions are provided by controllers rather than multiple vendor-specific devices and protocols \cite{onf_2012}.

\subsection{SDN Architecture}
\label{subsec:4.3.1}

In SDN model, network devices simply become specialized elements in packet forwarding tasks, which are executed according to a set of rules established by one or more controllers in control layer. These rules generally are predefined routines in application layer. On the other hand, the controller is a device located on a remote network and it communicates with the other network devices through a secure connection using a set of standardized commands \cite{jarraya_2014,muro_2016,xia_2015}.

As Fig.~\ref{ch4_fig:4} shows, in the reference model for SDN, the lower layer or infrastructure layer is where switching and routing devices (switches and routers) as well as other network devices are located. The control layer is made up of the SDN controller and it is separated from physical devices. Application Layer includes services or software applications that allow network managers or users executed specific tasks (such as bandwidth estimation and dynamic load balancing) over  the entire network.

\begin{center}
\includegraphics[scale=0.9]{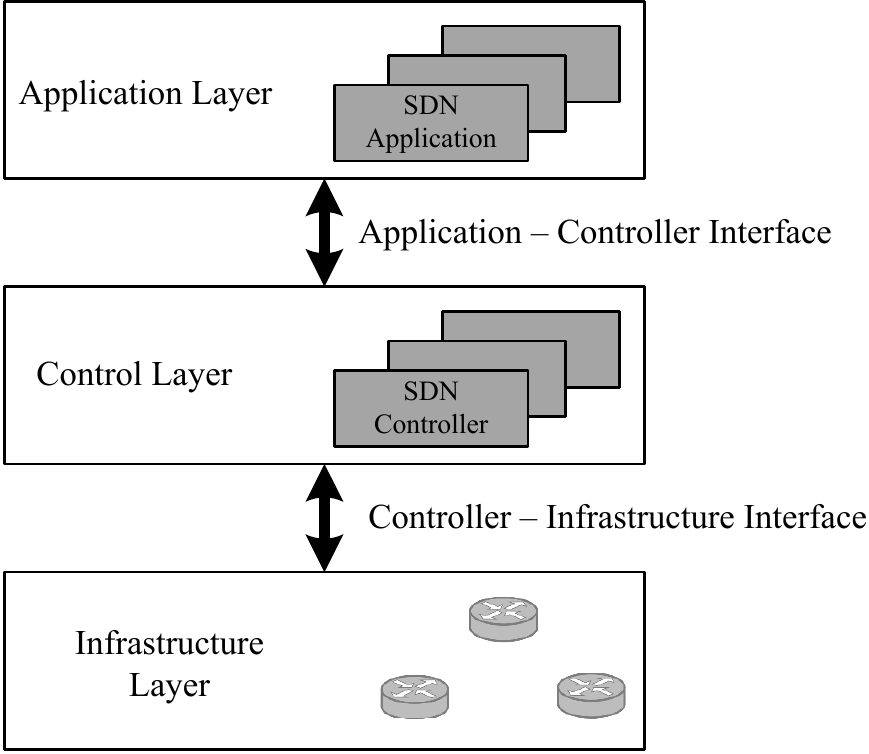}
\figcaption{Reference Model for SDNs}
\label{ch4_fig:4}       % Give a unique label
\end{center}

%\begin{figure}[h!]
%\centering
%\includegraphics[scale=0.7]{Figures/ch_4_fig_4_Reference_Model.eps}
%\caption{Reference Model for SDNs}
%\label{ch4_fig:4}       % Give a unique label
%\end{figure}

\section{Adaptive Video Transmission}
\label{sec:4.4}

Basically, video services are configured according to two main techniques: conventional video streaming and adaptive streaming. Conventional techniques include simple progressive download and real time streaming methods. In the progressive download technique, users request video content which is downloaded progressively into a local memory (buffer). According as the buffer has enough data the video starts to play. When playback rate overpasses the download rate, playback is stopped until more data is downloaded (see Fig.~\ref{ch4_fig:5}). Using a buffer some problems are solved, such as packet delay and the audio-video synchronization. Another conventional technique es the real time streaming, which consists on the establishment of a session between the video provider and users using a specialized network protocol \cite{castellanos_quality_2015}.

\begin{center}
\includegraphics[scale=0.6]{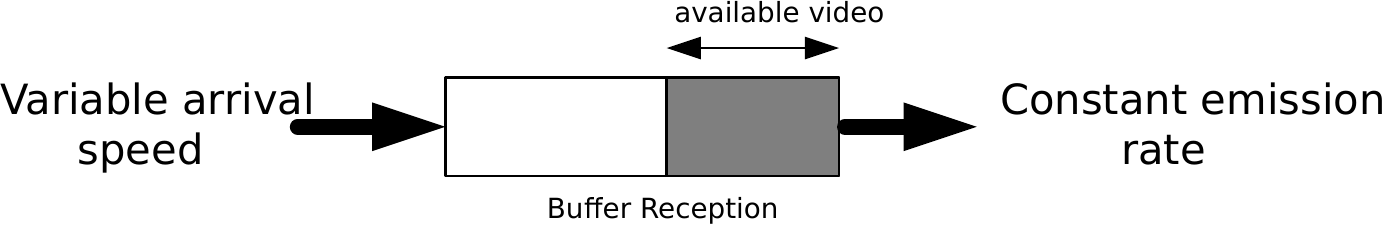}
\figcaption{Progressive download technique}
\label{ch4_fig:5}       % Give a unique label
\end{center}

%\begin{figure}[h!]
%\centering
%\includegraphics[scale=0.6]{Figures/ch_4_fig_5_Progressive_download.eps}
%\caption{Progressive download technique}
%\label{ch4_fig:5}       % Give a unique label
%\end{figure}

However, conventional methods for video streaming are not flexible techniques since video content is delivered to the client using a constant bit rate over a fluctuating  connection. Therefore, if a communication link is suffering congestion, quality of the video is significantly degraded, as preliminary studies have revealed \cite{castellanos_quality_2015,gonzalez_2016}.

Some alternatives for conventional video delivery methods are the adaptive streaming techniques. These techniques allow video server to adapt the video quality that is provided to the client according to the available network resources. Therefore, algorithms for estimation of available resources, for example in terms of available bandwidth, have to be implemented. Some estimation mechanisms proposed can be  consulted in \cite{castellanos_available_2019,chaud_2015}. 

There are three techniques for adaptive video transmission proposed to reduce the problem that occurs with traditional transmission. One technique is the transcoding, which is a process converts a video to another using specific parameters during encode process in order to satisfy a certain bit rate \cite{gonzalez_2016}. This technique presents some problems due to permanent fluctuations in communication links which implies many adaptations of video flow. Therefore, a high consumption of resources in video server is carried out.

The second adaptive video streaming method is the flow switching \cite{bing_2015}. In this method a video is encoded several times obtaining many versions of the same video, each with different bit rates. Fig.~\ref{ch4_fig:6} shows an example of this method, where each video version is also divided into chunks of the same duration. A software program has to dynamically choose the video version that matches the bandwidth available in the network. Although this technique minimizes processing costs, storage and transmission requirements must be considered because the same video content is encoded several times \cite{gonzalez_2016}. An example of this technique is the MPEG-DASH standard. This standard defines that the client has exclusively the control of the adaptive streaming process \cite{bing_2015}. The DASH client is responsible for requesting the server information required to carry out the video transmission. For this, videos must be encoded using the DASH standard. When a video is encoded in DASH, a file called manifest (or mpd) is generated, which must be stored on the server. The different representations of the video are also stored there. These representations are different versions of the original video, each version with different quality. These representations of the video are subdivided into segments (chunks) of a short duration. In this way, if network resources are very restrictive, DASH client can request the transmission of segments of the lowest quality representation to server.  Segments of the representation of best quality can be transmitted as more resources are available. A description of the video transmission using DASH is given in Fig.~\ref{ch4_fig:7}. As can be seen in this figure, several representations of the same video are stored on the HTTP server. Each representation has a specific bit rate and is composed of several segments. The mpd file is sent from server to client, so that it knows the configuration of the different representations. Therefore, client requests the suitable representation according to its available resources.

\begin{center}
\includegraphics[scale=1.0]{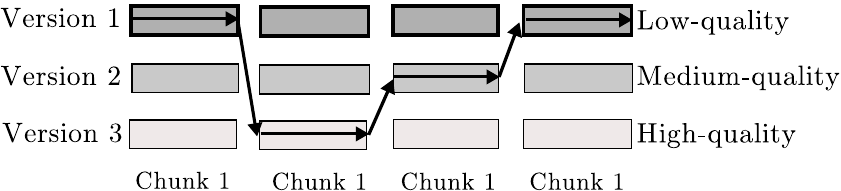}
\figcaption{Flow switching method}
\label{ch4_fig:6}       % Give a unique label
\end{center}

%\begin{figure}[h!]
%\centering
%\includegraphics[scale=1.0]{Figures/Ch_4_Fig6_flow_switching.eps}
%\caption{Flow switching method}
%\label{ch4_fig:6}       % Give a unique label
%\end{figure}

\begin{center}
\includegraphics[scale=0.7]{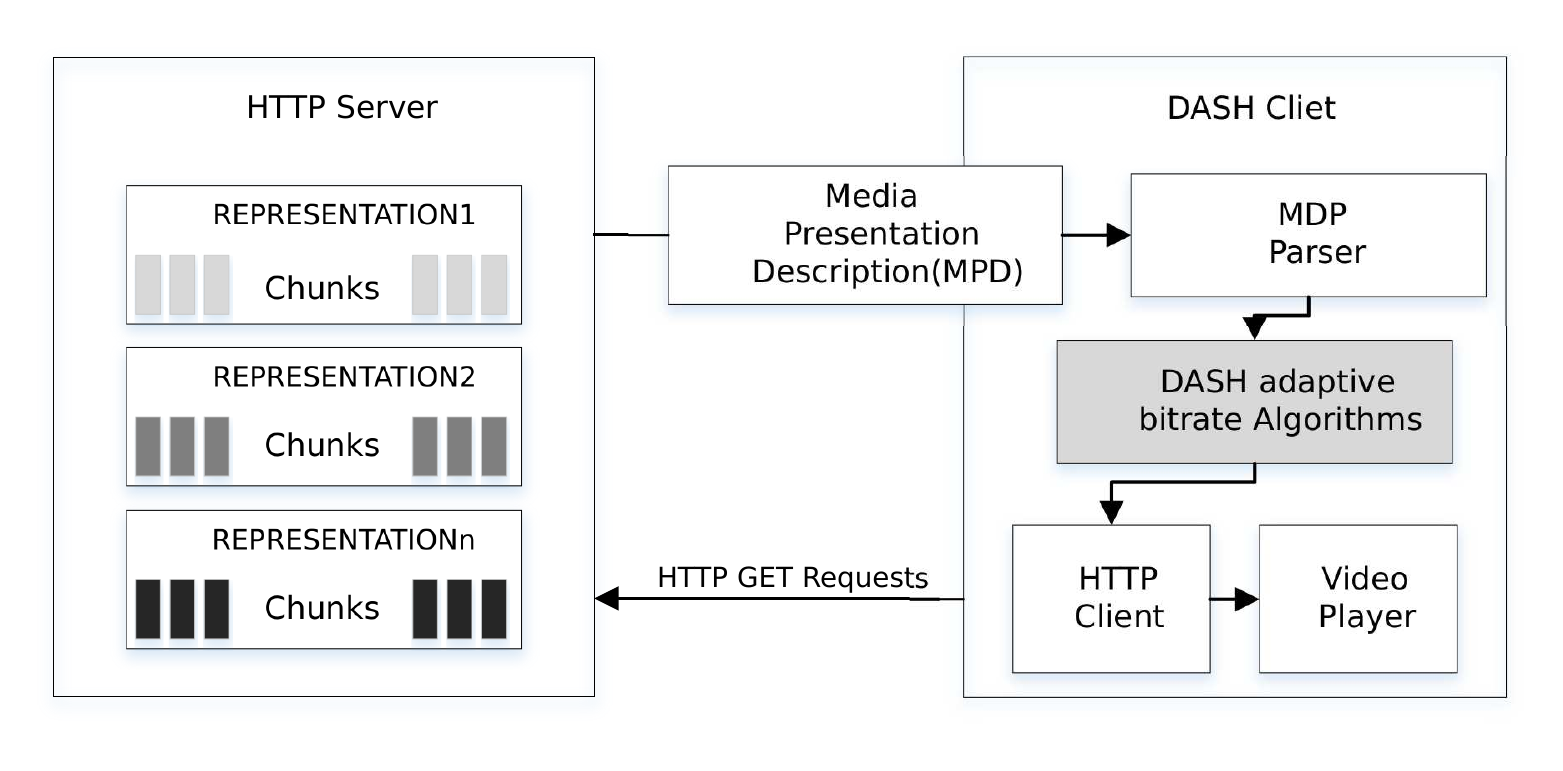}
\figcaption{DASH Video Transmission \protect\cite{kua_2017}}
\label{ch4_fig:7}       % Give a unique label
\end{center}

%\begin{figure}[h!]
%\centering
%\includegraphics[scale=0.7]{Figures/ch_4_fig_7_DASH_Video_Transmission.eps}
%\caption{DASH Video Transmission \cite{kua_2017}}
%\label{ch4_fig:7}       % Give a unique label
%\end{figure}

The third technique for adaptive video streaming uses Scalable Video Coding, which has attracted notable interest by researchers in order to improve video delivery over hybrid networks \cite{castellanos_improving_2018} and wireless networks \cite{castellanos_svceval-ra_2017}. The most recent standard for scalable video coding is known as SHVC (Scalable High Efficiency Video Coding). A video encoded with SHVC standard has a layered structure where the layers store information about different temporal, spatial or quality representations. The base layer corresponds to the more basic representation of the video. To this basic version of the video can be added several layers, known as enhancement layers. One or more enhancement layers increase the frame size, frame rate or video quality. The layered scheme of SHVC allow video server to send only layer data that can be supported by communication network. Fig.~\ref{ch4_fig:8} illustrates an example of the basic operation of the scalable transmission of a video containing three layers, each layer with different quality.

\begin{center}
\includegraphics[scale=0.43]{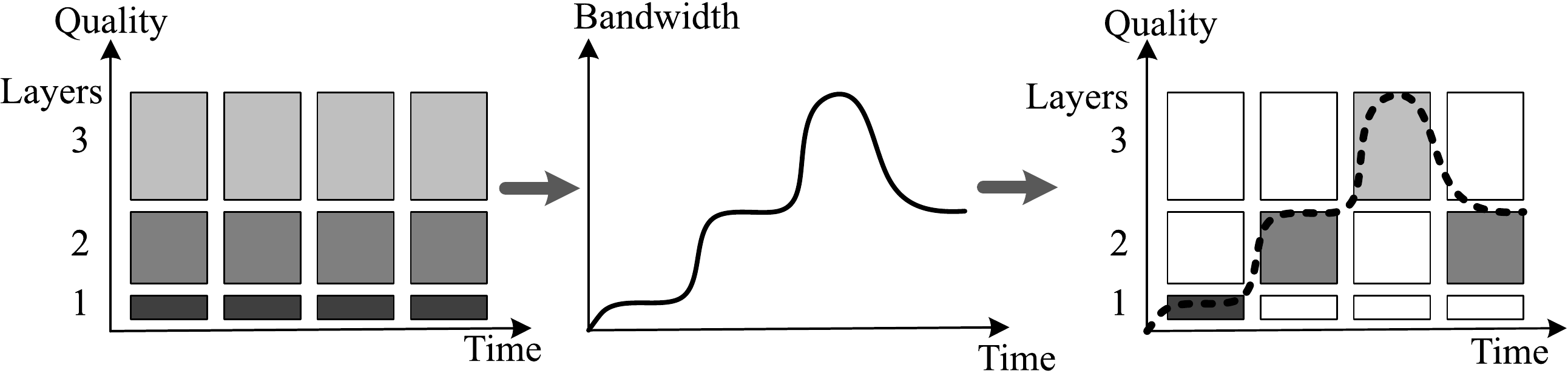}
\figcaption{Scalable Video Transmission}
\label{ch4_fig:8}       % Give a unique label
\end{center}

%\begin{figure}[ht!]
%\centering
%\includegraphics[scale=0.43]{Figures/Ch_4_Fig8_scalable.eps}
%\caption{Scalable Video Transmission}
%\label{ch4_fig:8}       % Give a unique label
%\end{figure}

\section{SHVC and DASH: tools and comparative evaluation}
\label{sec:4.5}
After understanding the operation and advantages of the DASH and SHVC standards in adapting video streams during a transmission, results of evaluating video transmission with these two methods are now described. This study has two objectives. On the one hand, to identify the main elements that make up an SDN and those components required to perform video traffic evaluations on this type of networks. And, on the other hand, to carry out a performance evaluation of adaptive video transmission over software-defined networks using DASH and SHVC techniques.

In experiments, a same video was transmitted using the two adaptive techniques (DASH and SHVC) in order to analyze which of them have better performance. In addition to adaptive streams, a video encoded in H.265 (non-adaptive) was also transmitted in order to determine the advantage of adaptive techniques respect to traditional streaming method.

The network scenario used in this evaluation is the NSFNET-14 network (National Science Foundation's Network). This network corresponds to a national optical network with 14 nodes, located in the United States and destined for research and education.

\subsection{Proposed methodology}
\label{subsec:4.5.1}
The methodology followed in the experiments described in this chapter is similar to the methodology of the previous experiments. 

\begin{center}
\includegraphics[scale=0.9]{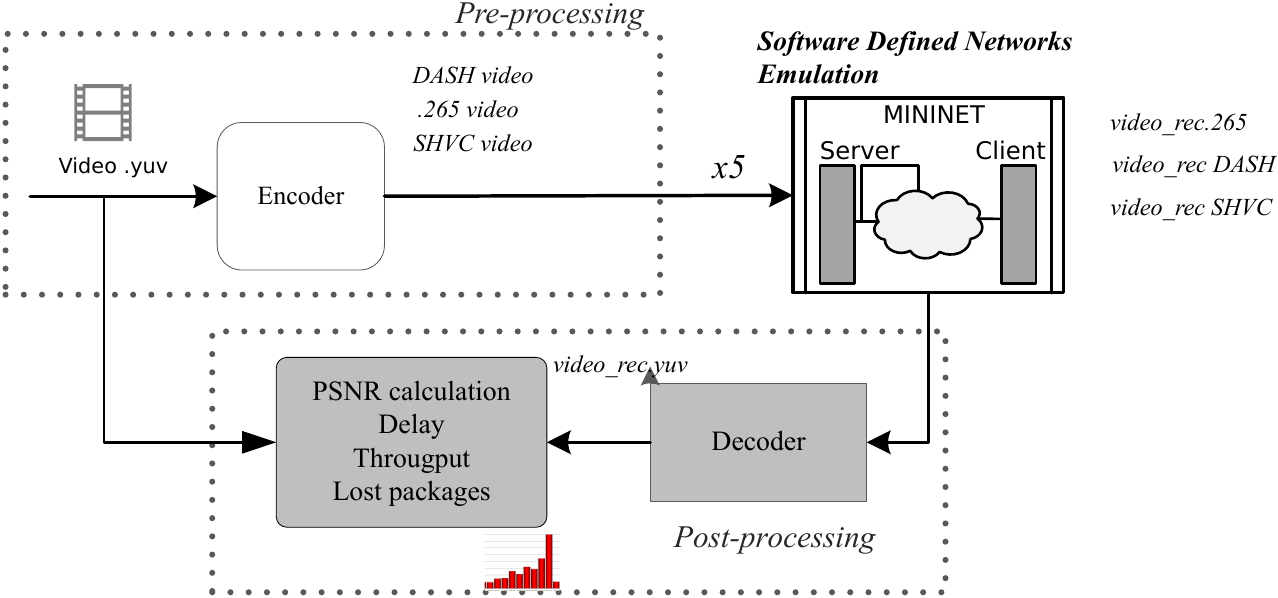}
\figcaption{Methodology for the development of the experiments}
\label{ch4_fig:9}       % Give a unique label
\end{center}

%\begin{figure}[ht!]
%\centering
%\includegraphics[scale=0.8]{Figures/Ch_4_Fig9_methodology.eps}
%\caption{Methodology for the development of the experiments}
%\label{ch4_fig:9}       % Give a unique label
%\end{figure}

Basically, there is a first preprocessing stage, in which the original video is encoded. In these experiments a testing video of 1200 frames was generated concatenating the well-known video sequences “mobile.yuv” with CIF resolution (352x288 pixeles) (see Fig.~\ref{ch4_fig:10}). This video was encoded with three tools in order to obtain three videos with different formats: H.265, DASH and SHVC. The H.265 coding was performed with the HM encoder; DASH video was generated with HM encoder to obtain an H.265 stream and then with MP4Box program to generate the DASH segments. Finally, SHM software was used to obtain the encoded video in SHVC. In the case of the H265 encoding, the video has a level of compression (CRF, Constant Rate Factor) of 20, in DASH there are three representations with three different levels of compression CRF of 20, 30 and 40, each with eight segments of 10 seconds of duration. SHVC Video has two layers, one with CRF of 20 and another of 40. It is important to remark that SHM encoder (official encoder for SHVC) allows maximum coding with 2 layers, therefore it was not possible to obtain a video with more scalability layers.

\begin{center}
\includegraphics[scale=0.3]{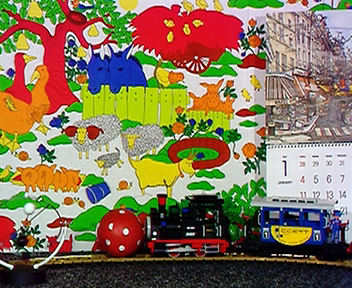}
\figcaption{Example frame of mobile.yuv}
\label{ch4_fig:10}       % Give a unique label
\end{center}

%\begin{figure}[ht!]
%\centering
%\includegraphics[scale=0.5]{Figures/Ch_4_Fig10_example.png}
%\caption{Example frame of mobile.yuv}
%\label{ch4_fig:10}       % Give a unique label
%\end{figure}

The second step of the methodology is the emulation of the network scenario and the independent transmission for each of the three videos. Thus, a video was received for each case. During video transmissions, traffic statistics of the video packets were recorded, which allowed that network performance be analyzed later. The network scenario, called NSFNET-14, was implemented in MININET, and consists of 14 nodes (see Fig.~\ref{ch4_fig:11}). The client and the server are located at opposite ends of the network, with host h1 being the client and h2 the server, as shown in Fig.~\ref{ch4_fig:11}. Bandwidth fluctuations were programmed as follows. Initially capacity of the client link is 400Kbps, then increases to 1200Kbps, then increases again to 2000Kbps and finally decreases to 1200Kbps. These bandwidth variations were made in the following instants of time: 3 seconds after the emulation starts, 14 seconds and 25 seconds, respectively. In addition, for this topology it was necessary to establish and configure routes (bidirectional routes) between client node and the server. This route management was done using the OpenFlow protocol.

\begin{center}
\includegraphics[scale=1.0]{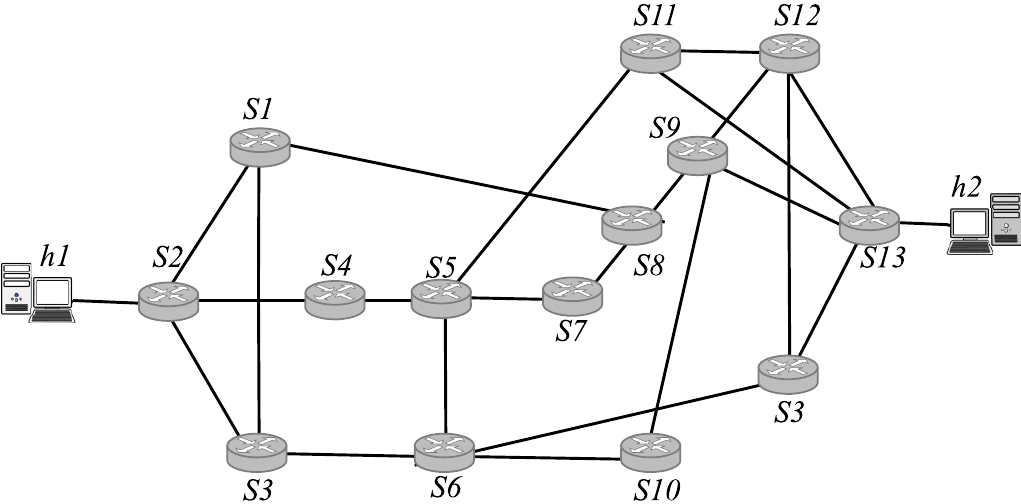}
\figcaption{NSFNET-14 network topology}
\label{ch4_fig:11}       % Give a unique label
\end{center}

%\begin{figure}[ht!]
%\centering
%\includegraphics[scale=1.0]{Figures/Ch_4_Fig11_NSFNET14.eps}
%\caption{NSFNET-14 network topology}
%\label{ch4_fig:11}       % Give a unique label
%\end{figure}

The last methodology stage was the post-processing. In this stage the received videos were decoded to YUV format to compare them with the original RAW video in order to measure the video quality in terms of PSNR (Peak Signal Noise Rate). PSNR is an objective metric used to estimate the image quality of a video in comparison with its original version. Therefore, the calculation of the PSNR gives us a clear quantification of the distortion suffered by the video. This metric is one of the most used to evaluate the quality of a video after its transmission over a network. Equation (1) shows the definition of the PSNR \cite{bing_2015}.

\begin{equation}
    PSNR=10 \cdot log_{10}( \dfrac{255^2}{MSE}) \ \ \ \ \textup{where,}
\end{equation}

\begin{equation}
    MSE=\frac{1}{N \cdot M} \sum^{M-1}_{m=0} \sum^{N-1}_{n=0} |I_{org} (m,n) - I_{rec} (m,n)|^2
\end{equation}

Where, Iorg is the original frame and Irec is the frame received after transmission; M, N is the size of the frame and MSE is the mean square error. According to (2), the PSNR can be understood as the comparison pixel by pixel, of each frame contained in the two videos (original and received). Therefore, the PSNR metric is a value that indicates the distortion experienced by received video regarding the original.

Other network performance metrics as delay, throughput and percentage of lost packets were calculated. To obtain that metrics a traffic capture was performed during video transmission. This capture of traffic was carried out using TCPDUMP software. However, in order to automate the data processing of traffic records, a Python program was implemented. This program integrates other scripts programmed in different programming languages such as Perl and Awk, as well as native Linux applications such as Bash, Grep, Sed, Join, Sort and Gnuplot. For additional information on the source code and operation of the program, can be consulted \cite{bustos_2019}.

\subsection{Simulation results}
\label{subsec:4.5.2}

In this section we present the results obtained from the network emulation and transmission of the three videos. It is important to note that the transmission of H.265, DASH and SHVC videos, was performed five times, in order to obtain statistically significant results. Therefore, each data points in figures represents an average of at least five runs with identical traffic.

Quality of the received videos is shown in Fig.~\ref{ch4_fig:12}, where PSNR of the three video formats is compared. The average PSNR for each video was also calculated, obtaining a PSNR of 11.35 dB, 27.67 dB and 31.01 dB for H265, SHVC and DASH videos, respectively.

\begin{center}
\includegraphics[scale=0.6]{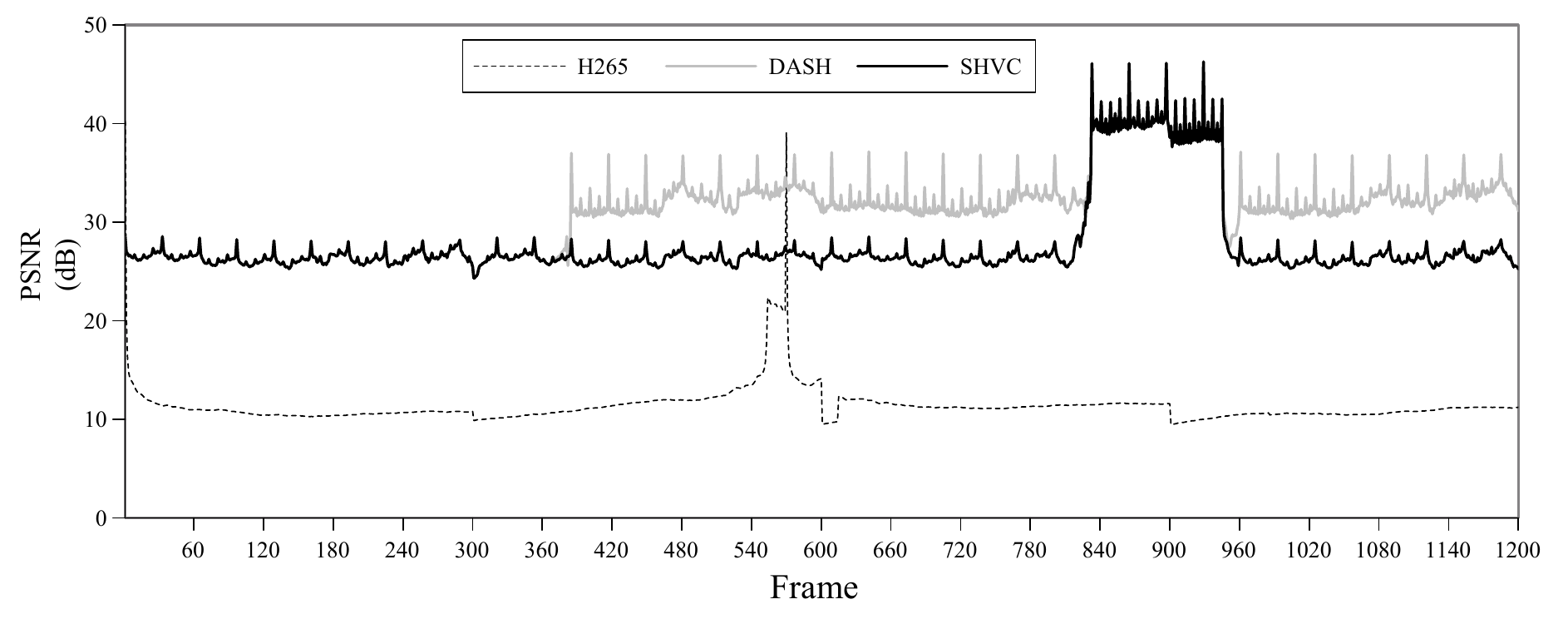}
\figcaption{Average PSNR of transmissions made on the NSFNET-14 topology.}
\label{ch4_fig:12}       % Give a unique label
\end{center}

%\begin{figure}[ht!]
%\centering
%\includegraphics[scale=0.6]{Figures/Ch_4_Fig12_averagePSNR.eps}
%\caption{Average PSNR of transmissions made on the NSFNET-14 topology.}
%\label{ch4_fig:12}       % Give a unique label
%\end{figure}

The PSNR curves allow to identify quality changes caused by the variation of the available bandwidth. Also, in Fig.~\ref{ch4_fig:12} the adaptation of the SHVC and DASH flows can be observed. On the contrary, using H265 encoding a very low PSNR (low quality) was obtained. The adaptability of DASH and SHVC streams allows user to obtain a higher video quality compared to quality of the H.265 stream (non-adaptive). The PSNR difference of H.265 in relation to DASH is 19.66 dB, and 16.32 dB in relation to SHVC, which shows that adaptive flows provide a significant improvement in quality.

In order to show that PSNR differences are reflected in video quality perceived by user, Fig.~\ref{ch4_fig:13} shows the image quality of a frame as an example. Fig.~\ref{ch4_fig:13} (a) shows the frame 510 of video encoded in H.265 and Fig.~\ref{ch4_fig:13}(b) shows the heat map resulting from the comparison with the same frame of the original video. Similarly, in Fig.~\ref{ch4_fig:13}(c) and Fig.~\ref{ch4_fig:13}(e) illustrate frames 510 of SHVC and DASH videos, respectively. The respective heat maps for these flows are also shown in Fig.~\ref{ch4_fig:13}(d) and Fig.~\ref{ch4_fig:13}(f). Red pixels in heat maps represent areas with high information losses (i.e. a low PSNR), while blue pixels represent best image quality. In Figure 13 below each heat map the PSNR value is shown. In this case, the frame with the lowest PSNR is correspond to H.265 video.

\begin{center}
\includegraphics[scale=0.6]{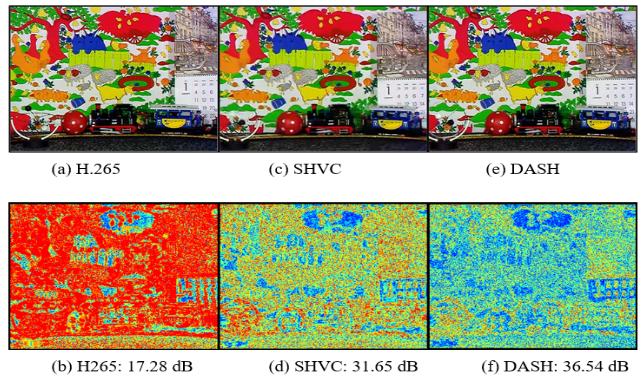}
\figcaption{Image quality of frame 510 for the three types of video.}
\label{ch4_fig:13}       % Give a unique label
\end{center}

%\begin{figure}[ht!]
%\centering
%\includegraphics[scale=0.6]{Figures/Ch_4_Fig13_image.png}
%\caption{Image quality of frame 510 for the three types of video.}
%\label{ch4_fig:13}       % Give a unique label
%\end{figure}

Network traffic behavior was also assessed throughout the measurement of throughput metric. In Fig.~\ref{ch4_fig:14} is depicted the variation of the total network throughput achieved while videos were transmitted from server to client node. Also, the variation of link capacity is shown in that Figure. The average values of the throughput for each case were: 1.22 Mbps, 243.5 Kbps and 377.5 Kbps, for H265, SHVC and DASH, respectively.

By analyzing curves in Fig.~\ref{ch4_fig:14}, it can be observed that in the cases of SHVC and DASH network is not saturated, since only packets belonging SHVC layer or DASH representation that can be supported, were transmitted over the network. Regarding H.265 video, server always tries to transmit video packets at highest quality, even when link bandwidth is more restrictive. Therefore, network is prone to congestion degrading the overall flow transmissions.

\begin{center}
\includegraphics[scale=0.6]{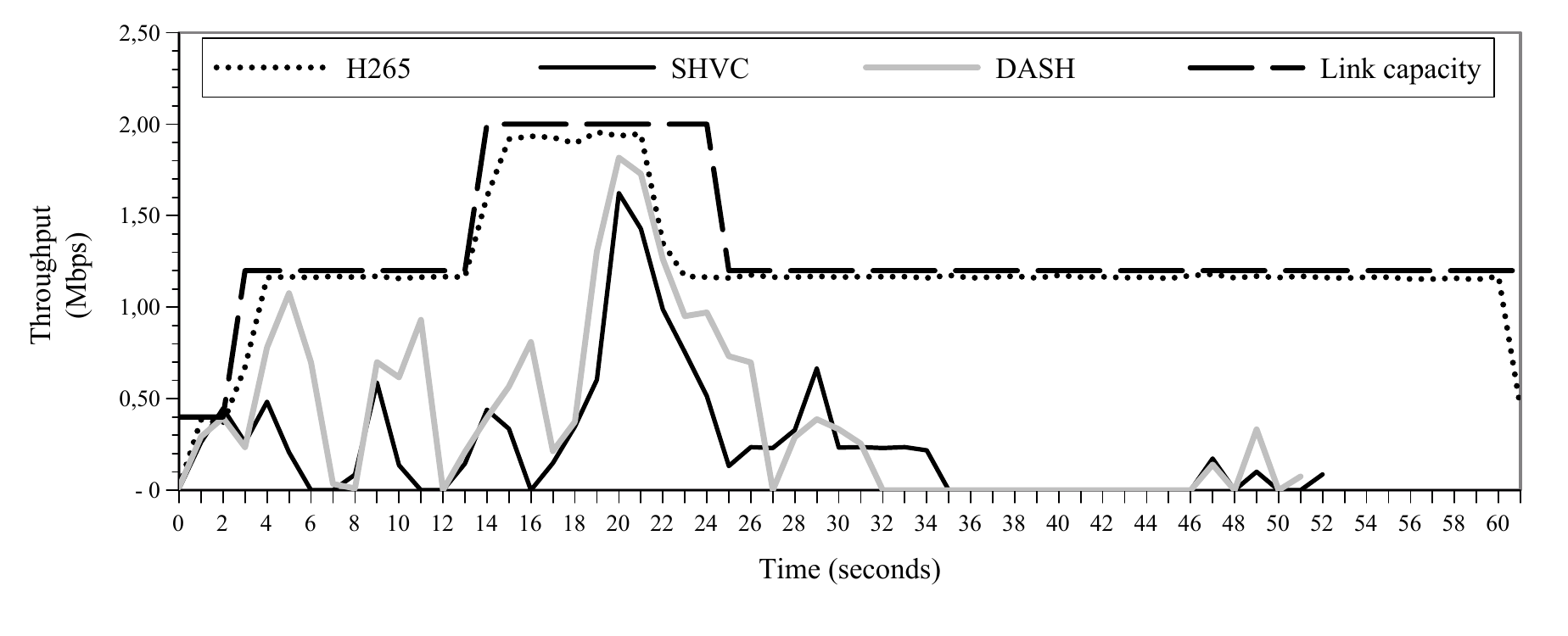}
\figcaption{Throughput achieved for each video transmission.}
\label{ch4_fig:14}       % Give a unique label
\end{center}

%\begin{figure}[ht!]
%\centering
%\includegraphics[scale=0.6]{Figures/Ch_4_Fig14_averageThroughput.eps}
%\caption{Throughput achieved for each video transmission.}
%\label{ch4_fig:14}       % Give a unique label
%\end{figure}

Network congestion for the H.625 flow can be confirmed if the end to end delay is analyzed. Packet delay values for each video stream are presented in Fig.~\ref{ch4_fig:15}. Average values for the packet delay were calculated, obtaining 1.3417, 0.0972 and 0.0995 seconds, for H265, SHVC and DASH, respectively. As can be observed, packets of adaptive flows presented a low delay. This means that network congestion was avoided. However, for H265 flow, a high delay affected its packets and this an indicator of network congestion. These high delay values influenced the perceived quality of the H.265 video since a smooth reproduction of the video was not obtained.

\begin{center}
\includegraphics[scale=0.6]{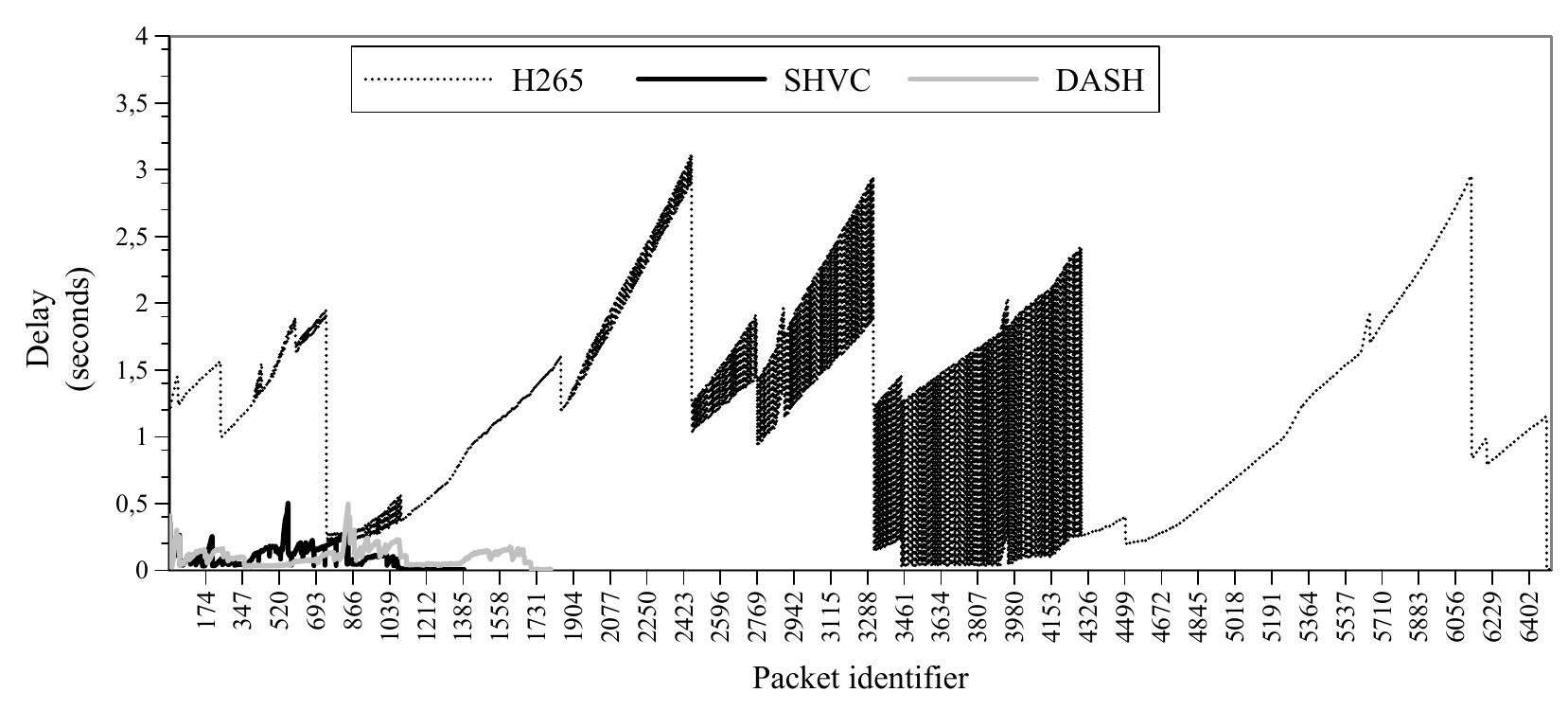}
\figcaption{Average end to end delay for each video flow.}
\label{ch4_fig:15}       % Give a unique label
\end{center}

%\begin{figure}[ht!]
%\centering
%\includegraphics[scale=0.6]{Figures/Ch_4_Fig15_averageDelay.eps}
%\caption{Average end to end delay for each video flow.}
%\label{ch4_fig:15}       % Give a unique label
%\end{figure}

On the other hand, throughput achieved for each video flows can be confirmed counting the number of sent packets during video transmissions. Table~\ref{ch4_tab:1} shows that using SHVC fewer packets were transmitted, only 1.155 packets. DASH flows used 1718 packets and with the highest amount of sent packets, the H265 video. The last metric obtained related to the network performance was the packet loss, which was represented in terms of percentage. This metric considers the packets sent by video server and those received by client host, calculated values are shown in Table~\ref{ch4_tab:1} and depicted in Fig.~\ref{ch4_fig:16}.

\begin{table}[ht!]
    \centering
\caption{Packet statistics in video transmissions}
\label{ch4_tab:1}
    \begin{tabular}{|c|c|c|c|c|}
         \hline
 & Sent Packets & Received Packets & Lost Packets & Percentage of lost packets \\
\hline
H265 & 6.421,6 & 6.329,0 & 92,9 & 1.44\% \\
\hline
SHVC & 1.155 & 1.142,2 & 12,8 & 1,1\% \\
\hline
DASH & 1.718,2 & 1.703,2 & 15 & 0.87\%\\
\hline
    \end{tabular}
\end{table}

\begin{center}
\includegraphics[scale=0.6]{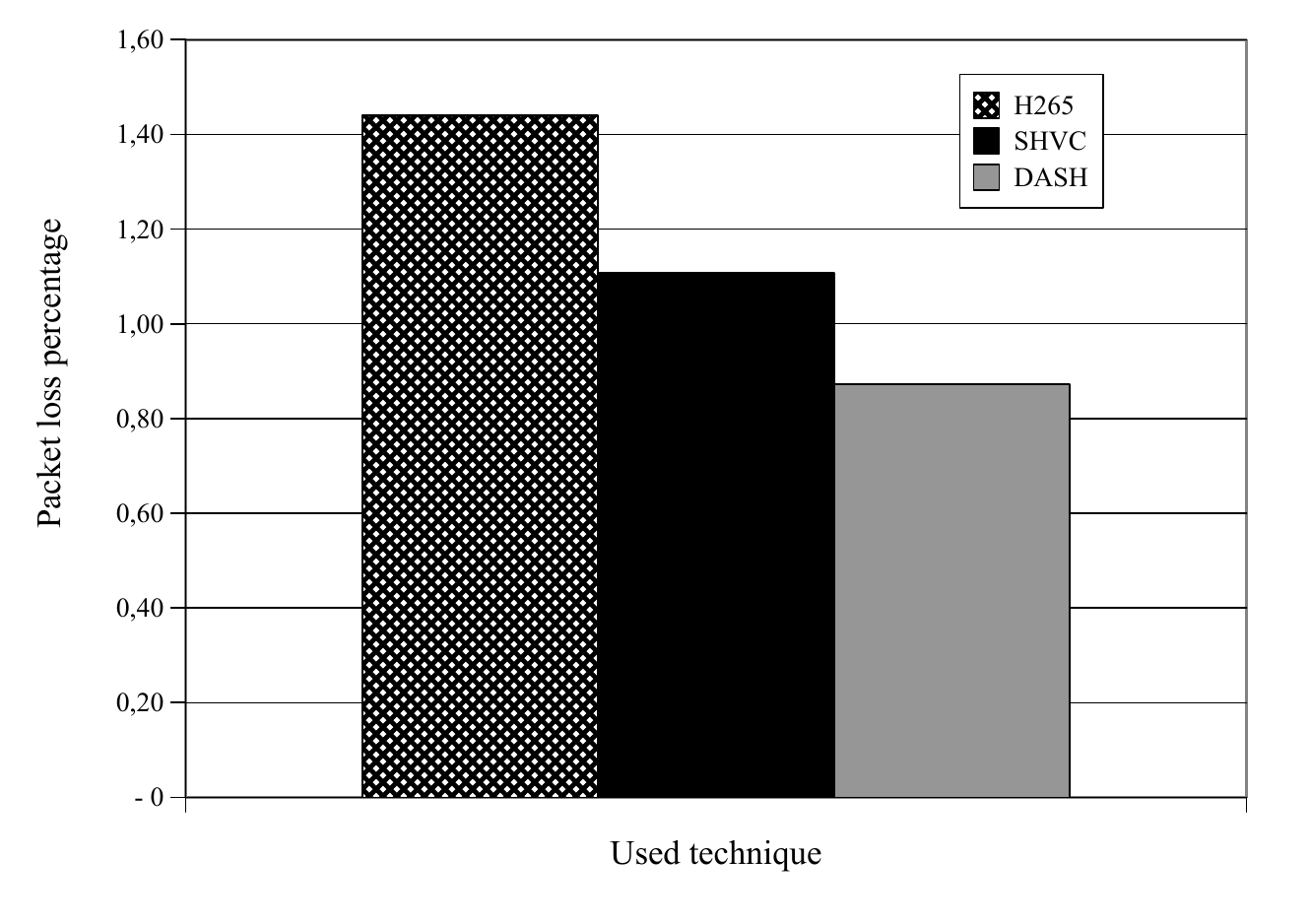}
\figcaption{Percentage of lost packets.}
\label{ch4_fig:16}       % Give a unique label
\end{center}

%\begin{figure}[ht!]
%\centering
%\includegraphics[scale=0.6]{Figures/Ch_4_Fig16_percentageLoss.eps}
%\caption{Percentage of lost packets.}
%\label{ch4_fig:16}       % Give a unique label
%\end{figure}

Although the number of lost packets is low for the three streams, the H.265 stream experienced the highest amount of packets loss. A low percentage of lost packets and a low delay (as in the case of DASH and SHVC) means a better adaptation of the traffic flow to the maximum link capacity. 

\section{Conclusions}
\label{sec:4.6}

Initially in this chapter we present important forecasts about the massive explosion of the video services and the consequent increase in the real time traffic on the current network infrastructures. Then, the main aspects concerning video transmission over software defined networks were analyzed. Particularly, techniques for adaptive video streaming were presented since these techniques are the most suitable methods for multimedia content delivery over networks with limited available resources. 

One of the most important contributions of this chapter is the methodology and  the tools proposed for developing video transmission experiments over SDNs. Using this tool set we carried out a study on adaptive video streaming with three different encoding standards: H.265, SHVC and DASH. These standards are the most recent for video coding and they are the encoding standards that will be used in near future for developing the next generation of UHD (Ultra High Definition) video services.

Regarding video transmission evaluation on SDNs, adaptive techniques like DASH and SHVC outperform H.265 in terms of quality of received video, throughput and video packet delay. Particularly, DASH proved to be an efficient adaptation technique since it allowed to obtain not only higher quality videos, but also experienced low delay and low packet losses.

Using SHVC good results also were obtained. Although the ability of SHVC to effectively adapt to the capacity of the link could be checked, its current limitation of only being able to add one improvement layer makes it less flexible. The great advantage of SHVC is that only one version of the video is stored on server, while several versions of the same video must be stored with DASH, which significantly increases the disk space used.

The worst results obtained were reached by transmitting the video encoded only with H.265 standard, without any adaptive technique. Although more video information is sent with H265, the lack of adaptation to variations in available bandwidth of the data network causes network congestion. Therefore, the delay and packet losses significantly increase, obtaining a received video with very low quality.

On the contrary, with adaptive techniques, although not all information contained in the original video is sent, a better quality of experience is obtained by user. This fact that can be demonstrated evaluating the PSNR.

On the other hand, throughput measurements show that adaptive techniques make more efficient use of bandwidth, preventing congestion network and avoiding affect other simultaneous traffic flows. This situation also allows delay and packet loss to be noticeably smaller, compared to transmissions where adaptive video techniques are not used.
Finally, we can conclude that adaptive techniques are absolutely essential for the transmission of current high-performance video services, both in conventional networks and in the new model defined by the SDN.

\section*{Acknowledgements}
 This work is supported by the Universidad de San Buenaventura under the project “PERSEO: Plataforma para Evaluar los Sistemas de vidEO de las aplicaciones de próxima generación” (CBI 013-008)

\backmatter%%%%%%%%%%%%%%%%%%%%%%%%%%%%%%%%%%%%%%%%%%%%%%%%%%%%%%%
%\chapter*{}
%\addcontentsline{toc}{chapter}{Bibliography}

\bibliographystyle{apacite}
\bibliography{
biblio_ch_4_}
\printindex

%%%%%%%%%%%%%%%%%%%%%%%%%%%%%%%%%%%%%%%%%%%%%%%%%%%%%%%%%%%%%%%%%%%%%%

\end{document}